\documentclass[11pt]{article}
\pdfoutput=1
\usepackage{amsfonts,amsmath}
\usepackage{amssymb}
\usepackage{fancyhdr}
\usepackage{slashed}
\usepackage{graphicx}
\usepackage{subfigure}
\usepackage{color}

\usepackage{hyperref}
\usepackage[utf8]{inputenc}
\usepackage[titletoc]{appendix}

\def\hybrid{\topmargin -20pt    \oddsidemargin 0pt
        \headheight 0pt \headsep 0pt
        \textwidth 6.25in       
        \textheight 9.25in       
        \marginparwidth .875in
        \parskip 5pt plus 1pt   \jot = 1.5ex}

\hybrid

\def\baselinestretch{1.2}

\catcode`\@=11

\def\marginnote#1{}
\newcount\hour
\newcount\minute
\newtoks\amorpm
\hour=\time\divide\hour by60
\minute=\time{\multiply\hour by60 \global\advance\minute by-\hour}
\edef\standardtime{{\ifnum\hour<12 \global\amorpm={am}%
        \else\global\amorpm={pm}\advance\hour by-12 \fi
        \ifnum\hour=0 \hour=12 \fi
        \number\hour:\ifnum\minute<10 0\fi\number\minute\the\amorpm}}
\edef\militarytime{\number\hour:\ifnum\minute<10 0\fi\number\minute}

\def\draftlabel#1{{\@bsphack\if@filesw {\let\thepage\relax
   \xdef\@gtempa{\write\@auxout{\string
      \newlabel{#1}{{\@currentlabel}{\thepage}}}}}\@gtempa
   \if@nobreak \ifvmode\nobreak\fi\fi\fi\@esphack}
        \gdef\@eqnlabel{#1}}
\def\@eqnlabel{}
\def\@vacuum{}
\def\draftmarginnote#1{\marginpar{\raggedright\scriptsize\tt#1}}

\def\draft{\oddsidemargin -.5truein
        \def\@oddfoot{\sl preliminary draft \hfil
        \rm\thepage\hfil\sl\today\quad\militarytime}
        \let\@evenfoot\@oddfoot \overfullrule 3pt
        \let\label=\draftlabel
        \let\marginnote=\draftmarginnote
   \def\@eqnnum{(\theequation)\rlap{\kern\marginparsep\tt\@eqnlabel}%
\global\let\@eqnlabel\@vacuum}  }

\def\preprint{\twocolumn\sloppy\flushbottom\parindent 2em
        \leftmargini 2em\leftmarginv .5em\leftmarginvi .5em
        \oddsidemargin -.5in    \evensidemargin -.5in
        \columnsep .4in \footheight 0pt
        \textwidth 10.in        \topmargin  -.4in
        \headheight 12pt \topskip .4in
        \textheight 6.9in \footskip 0pt
        \def\@oddhead{\thepage\hfil\addtocounter{page}{1}\thepage}
        \let\@evenhead\@oddhead \def\@oddfoot{} \def\@evenfoot{} }

\def\numberbysection{\@addtoreset{equation}{section}
        \def\theequation{\thesection.\arabic{equation}}}

\def\underline#1{\relax\ifmmode\@@underline#1\else
        $\@@underline{\hbox{#1}}$\relax\fi}

\def\titlepage{\@restonecolfalse\if@twocolumn\@restonecoltrue\onecolumn
     \else \newpage \fi \thispagestyle{empty}\c@page\z@
        \def\thefootnote{\fnsymbol{footnote}} }

\def\endtitlepage{\if@restonecol\twocolumn \else \newpage \fi
        \def\thefootnote{\arabic{footnote}}
        \setcounter{footnote}{0}}  

\catcode`@=12
\relax

\def\figcap{\section*{Figure Captions\markboth
        {FIGURECAPTIONS}{FIGURECAPTIONS}}\list
        {Figure \arabic{enumi}:\hfill}{\settowidth\labelwidth{Figure
999:}
        \leftmargin\labelwidth
        \advance\leftmargin\labelsep\usecounter{enumi}}}
 \relax
\def\tablecap{\section*{Table Captions\markboth
        {TABLECAPTIONS}{TABLECAPTIONS}}\list
        {Table \arabic{enumi}:\hfill}{\settowidth\labelwidth{Table
999:}
        \leftmargin\labelwidth
        \advance\leftmargin\labelsep\usecounter{enumi}}}
 \relax
\def\reflist{\section*{References\markboth
        {REFLIST}{REFLIST}}\list
        {[\arabic{enumi}]\hfill}{\settowidth\labelwidth{[999]}
        \leftmargin\labelwidth
        \advance\leftmargin\labelsep\usecounter{enumi}}}
 \relax
\makeatletter
\newcounter{pubctr}
\def\publist{\@ifnextchar[{\@publist}{\@@publist}}
\def\@publist[#1]{\list
        {[\arabic{pubctr}]\hfill}{\settowidth\labelwidth{[999]}
        \leftmargin\labelwidth
        \advance\leftmargin\labelsep
        \@nmbrlisttrue\def\@listctr{pubctr}
        \setcounter{pubctr}{#1}\addtocounter{pubctr}{-1}}}
\def\@@publist{\list
        {[\arabic{pubctr}]\hfill}{\settowidth\labelwidth{[999]}
        \leftmargin\labelwidth
        \advance\leftmargin\labelsep
        \@nmbrlisttrue\def\@listctr{pubctr}}}
 \relax
\makeatother
\newskip\humongous \humongous=0pt plus 1000pt minus 1000pt

\newif\ifdtup

\relax

\def\be{\begin{equation}}
\def\ee{\end{equation}}
\def\ba{\begin{eqnarray}}
\def\ea{\end{eqnarray}}

\def\k{\kappa}
\def\r{\rho}
\def\a{\alpha}

\def\b{\beta}

\def\g{\gamma}

\def\d{\delta}
\def\D{\Delta}

\def\th{\theta}
\def\Th{\Theta}
\def\m{\mu}
\def\mn{{\mu \nu}}
\def\n{\nu}

\def\l{\lambda}

\def\cL{{\cal L}}

 \def\cK{{\cal K}}
 \def\cL{{\cal L}}
  \def\cO{{\cal O}}

\newcommand{\vev}[1]{{\left< {#1} \right>}}
\newcommand{\bra}[1]{{\left< {#1} \right|}}
\newcommand{\ket}[1]{{\left| {#1} \right>}}
\newcommand{\prt}[1]{{\left( {#1} \right)}}
\newcommand{\prtt}[1]{{\left[ {#1} \right]}}

\def\no{\noindent}

\def\IR{\relax{\rm I\kern-.18em R}}

\def\pp{\partial}

\newcommand{\ff}{\dfrac}

\def\IR{\relax{\rm I\kern-.18em R}}
\def\IL{\relax{\rm I\kern-.18em L}}

\def\inv{^{\raise.15ex\hbox{${\scriptscriptstyle -}$}\kern-.05em 1}}

\def\cL{{\cal L}}

\def\tr{{\rm tr}}

\def\bea{\begin{eqnarray}}
\def\eea{\end{eqnarray}}
\newcommand{\eq}[1]{(\ref{#1})}
\def\nn{\nonumber}

\newcommand{\la}[1]{\label{#1}}

\def\a{\alpha}      
\def\b{\beta}       
\def\g{\gamma}    
\def\d{\delta}  \def\D{\Delta}

\def\k{\kappa}
\def\l{\lambda} 
\def\m{\mu} \def\n{\nu}
\def\o{\omega}
 
\def\r{\rho}
  
\def\t{\tau}
\def\th{\theta} \def\Th{\Theta}

\def \dt {\dot{t}}

\def \dr {\dot{r}}

\def \dth {\dot{\theta}}

\def \dphi {\dot{\phi}}

\def \dx {\dot{x}}
\def \ddx { \ddot{x}}
\definecolor{markcolor2}{rgb}{1,0,0}

\definecolor{markcolor3}{rgb}{0,1,0}

\begin{document}

\renewcommand{\theequation}{\thesection.\arabic{equation}}
\csname @addtoreset\endcsname{equation}{section}

\newcommand{\beq}{\begin{equation}}
\newcommand{\eeq}[1]{\label{#1}\end{equation}}
\newcommand{\ber}{\begin{eqnarray}}
\newcommand{\eer}[1]{\label{#1}\end{eqnarray}}
\newcommand{\eqn}[1]{(\ref{#1})}
\begin{titlepage}

\begin{center}

~
\vskip .7 cm

{\Large
\bf  Quasinormal Modes and Universality

}

{\Large
\bf   of the Penrose Limit  of Black Hole Photon Rings
}

\vskip 0.6in

 {\bf D. Giataganas${}^{1,2,3}$ , A. Kehagias${}^{4}$, A. Riotto${}^{5}$}
 \vskip 0.1in
 {\em
 	${}^1$  Department of Physics, National Sun Yat-sen University,
 	Kaohsiung 80424, Taiwan\\
 	${}^2$  Center for Theoretical and Computational Physics,
 	National Sun Yat-sen University \\
 	${}^3$ Physics Division, National Center for Theoretical
 		Sciences,
 	Taipei 10617, Taiwan\\
  ${}^4$Physics Division, National Technical University of Athens, Athens, 15780, Greece\\
 	${}^5$ Department of Theoretical Physics and Gravitational Wave Science Center (GWSC) \\
24 quai E. Ansermet, CH-1211 Geneva 4, Switzerland
 \\\vskip .15in
 {\tt ~  dimitrios.giataganas@mail.nsysu.edu.tw, kehagias@central.ntua.gr, antonio.riotto@unige.ch
 }\\
 }

\vskip .1in
\end{center}

\vskip .4in

\centerline{\bf Abstract}
\noindent
We study the physics of photon rings in a wide range of axisymmetric black holes admitting a separable Hamilton-Jacobi equation for the geodesics. Utilizing the Killing-Yano tensor, we derive the Penrose limit of the black holes, which describes the physics near the photon ring. The obtained plane wave geometry is directly linked to the frequency matrix of the massless wave equation, as well as the instabilities and Lyapunov exponents of the null geodesics. Consequently,  the Lyapunov exponents and frequencies of the photon geodesics, along with the quasinormal modes, can be all extracted from a Hamiltonian in the Penrose limit plane wave metric. Additionally, we explore potential bounds on the Lyapunov exponent, the orbital and precession frequencies, in connection with the corresponding inverted harmonic oscillators and we discuss the possibility of photon rings serving as 
effective holographic horizons in a holographic duality framework for astrophysical black holes. Our formalism is applicable to spacetimes encompassing various types of black holes, including stationary ones like Kerr, Kerr-Newman, as well as static black holes such as Schwarzschild, Reissner-Nordström, among others.

\no
\end{titlepage}
\vfill
\eject


\noindent


\def\baselinestretch{1.2}
\baselineskip 19 pt
\noindent


\setcounter{equation}{0}

\section{Introduction}

 All black holes have an event horizon, the area of which is associated with the entropy of the black hole, naturally related to the number of all possible microstates of the black hole. A common question is where these microstates lie in the quantum dual description. The most natural guess is that this information lies on the event horizon or a few Planck lengths from the horizon. The black holes admit bound null geodesics comprising the photon ring, which we are now able to indirectly observe with black hole imaging \cite{EventHorizonTelescope:2019dse}, and we will be able to gather more information in the future with space very long baseline interferometry \cite{Chael:2021rjo,Johnson:2019ljv}. These unstable null geodesics are responsible for the observation of the photon ring by distant observers, and  contain significant information about certain properties of black holes.

Any infalling object to the black hole, before reaching its event horizon and exciting the black hole, crosses the photon shell. It interacts with the photons of the nearly bound geodesics around the black hole, which escape the sphere at a much later time after orbiting around it several times \cite{Cardoso:2021sip}. These photons are expected to be the last signals observed after the black hole returns to its initial state after the excitation of the infalling object. There lies the relation between the quasinormal modes (QNMs) in the geometrical optics approximation and the lightlike geodesics \cite{Ferrari:1984zz,Cardoso:2008bp,Yang:2012he,Hod:2009td,Dolan:2010wr}. Apart from the obvious significance of the photon sphere as a fundamental probe of the black hole, the above description qualifies the photon sphere as a part of the holographic dual of asymptotically flat black holes \cite{Hadar:2022xag}. As a result, in-depth studies on the photon sphere of black holes provide an exceptionally interesting field for gravitational physics, including observational physics and quantum-theoretical black hole physics.

A central step forward in these studies is the development of a systematic generic framework for the study of the photon rings\footnote{Here we use the term “photon ring”  with a more general meaning, like many other relevant works. The photon ring itself, usually when defined strictly is the ring image produced by near photon shell photons when they reach a telescope at infinity.} that depends on the general characteristics and symmetries of the black holes, and is readily applicable to a large class of interesting spacetimes. This is a viable approach since the connection of the fundamental properties of the black holes associated with the photon ring is sensitive mostly to certain detailed specifics of the black holes.

In this work, we consider the Penrose limit of a wide class of generic spacetimes on a null geodesic to obtain a plane wave metric, in order to study the photon ring and the associated properties. Aspects of this approach were introduced in \cite{Fransen:2023eqj}. Our framework is based on a series of fundamental properties of the Penrose limit. We argue that certain significant near photon shell physics is captured in a straightforward way by the Penrose limit. The Penrose plane wave metric can be thought of as the geometry that observers see with velocities close to the speed of light, where their clocks are calibrated in such a way that the affine parameter along the null geodesics remains invariant, making the space-time non-degenerate. It isolates physics of the neighborhood of the null geodesics, in our case, the photon rings we like to examine.

It is known that any spacetime has, as a limit, a plane wave metric. Plane waves come as special cases of the pp-waves limit, with a zero gauge field corresponding to the cross metric elements between the parallel and transverse directions, while the metric elements of the parallel direction have quadratic dependence on the transverse directions. The plane wave in Brinkmann coordinates is almost unique due to the limiting coordinate transformations that can be done while respecting the gauge conditions imposed and the form of the metric. Moreover, the equations of motion for the null geodesics in the light-cone gauge are those of a non-relativistic harmonic oscillator \cite{blau}. Then the plane waves possess at minimum a Heisenberg algebra, associated with the null Killing vector of the spacetime. This is a minimal symmetry of the Penrose limit regardless of the symmetries of the original metric, and the Heisenberg algebra does not reflect further symmetries of the original metric.
However, when extra symmetries of the original spacetime exist, they will generate more Killing vectors, and therefore the Heisenberg algebra is extended, for instance, to the harmonic oscillator algebra for homogeneous plane waves.  The plane waves we construct are by taking the Penrose limit of a spacetime for chosen null geodesics that comprise the photon ring. Then the Penrose limit describes the near photon ring regime and symmetries that emerge in the photon ring physics are revealed by the plane waves.

There is a crucial reason that makes the study of photon rings with Penrose limits compelling. The covariant information contained in the Penrose limit of the original metric is related to the tidal forces along the null geodesic. In particular, the plane wave metric elements determine the transverse null geodesic deviation matrix via the curvature tensor of the parallel frame, and is independent of the geodesic embeddings \cite{blau,Blau:2003dz, Blau:2004yi, Hawkings_ellis}. The geodesic deviation in the original metric and in its Penrose limit plane wave are identical \cite{Blau:2003dz, Blau:2004yi}. Therefore, the Lyapunov exponents of the nearly bound geodesics that comprise the photon ring can be directly read from the Penrose limit plane wave of the black hole under study once it is constructed. At the large frequency limit, the massless scalar equation in the original metric can be approximately solved with waves whose wavefronts propagate along the null geodesics. The solutions of the massless wave equation are, in the eikonal approximation, related to congruences of null ring geodesics. In the Penrose limit, this correspondence takes a different form. The massless wave equation in the Penrose plane wave has a frequency matrix directly related to the frequency matrix of the null congruences. The deviation of the null geodesics is identical in the original geometry and in the plane wave limit, as it covariantly preserves this information. Therefore, it naturally follows that the wave equation in the Penrose limit is related to the deviation of the null geodesics in the original spacetime without any further limit needed to be taken. In particular, the Lyapunov exponents characterizing the exponential divergence of the asymptotically bound geodesics in the near photon shell regime correspond to the exponential decay of the quasinormal modes at the Penrose plane wave limit, realizing geometrically the geometric optics approximation. We show that the full analysis boils down to producing the Penrose limit of the original spacetime black hole metric. Once this is done, all the details are directly readable from the plane wave metric.

The correspondence has some direct and fascinating consequences. For spacetimes that are solutions of the vacuum Einstein equations in the absence of null fluxes, have Ricci flat Penrose limits. The matrix $A$  representing the non-trivial metric elements of the Penrose limit, is proportional to the curvature tensor of the parallel frame along the photon ring geodesics, and it is traceless. This immediately implies that the transformed Schrödinger equations of the massless wave equations, corresponding for example to the radial and angular geodesic deviations, always result in two "mirroring" harmonic oscillators. One is a normal harmonic oscillator along a stable polar direction with a frequency matrix proportional to the negative diagonal element of $A$, and the other is an inverted harmonic oscillator along a radial unstable direction differing only in sign of the frequency term. This also implies that the Lyapunov exponent characterizing the unstable perturbations is related to the stable angular frequency of the original orbit. This is not the case for black hole solutions to the equations of motion with matter fields or fluxes along the null direction, since the traceless property is lost. There, we may still obtain two different Schrödinger equations corresponding to the normal and inverted harmonic oscillators, but their frequency matrix elements are not of the same norm.

Another aim of our work is to examine and motivate possible bounds on the Lyapunov exponents and the quasi-normal modes that generalize well-known quantum bounds \cite{Maldacena:2015waa}, which can be expressed as  $\lambda \lesssim \kappa$, where $\kappa$ is the surface gravity. It is known that certain types of particle geodesics around unstable points in the near-horizon regime of the Schwarzschild black hole can have Lyapunov exponents that saturate a classical bound of chaos, where the Lyapunov exponent is equal to the surface gravity \cite{Hashimoto:2016dfz}. However, it has been later proved that this bound is not universal, and in generic black hole spacetimes, it does not hold despite enforcing the energy conditions and the equations of motion constraints \cite{Giataganas:2021ghs}. Nevertheless, in the case of the photon ring geodesics, we can map the system to a quantum inverted harmonic oscillator, with a potential identified by the Penrose limit metric. 
The plane wave metric and the Lyapunov exponent can be expressed through the surface gravity of the photon ring of the black hole in a meaningful pattern. In fact, for Schwarzschild-de Sitter black holes in four dimensions, the Lyapunov exponent of the nearly bound null geodesics is equal to the surface gravity \cite{Cardoso:2008bp}. Here, we generalize such relations with the surface gravity for generic spacetimes and point out how such relations are modified in the presence of angular frequencies and charges.  Moreover, by conjecturing that the photon ring  acts as an effective holographic horizon for thermal quantum systems we study the existence of possible bounds on the Lyapunov exponents and the quasinormal mode properties. We introduce the term 'effective holographic horizon', as it defines the temperature experienced by the dual field theory. In the limit of large mass, we analyze the effective temperatures that satisfy the Lyapunov bounds. For the case of the Schwarzschild black hole, the effective temperature $T_{eff}$  that saturates a classical Lyapunov bound of the form $\l=2 \pi T_{eff}$ on the photon ring,  turns out to be equal to the Unruh temperature of the photon ring when treated as a Rindler horizon, while rotation and the presence of charges lead to an increase in the effective temperature in a way that is in agreement with expectations from quantum systems, supporting the idea that photon rings can be treated as effective holographic horizons.

To this end we suggest an alternative upper bound for the Lyapunov exponents and the quasinormal modes spectrum, where the quantities are bounded by the inverse of entropy $\l\le \pi/(4 S)$, instead of the surface gravity. We show that this bound is satisfied universally on the photon rings for the black holes we consider in this paper.

Let us briefly review the main steps of our formalism. We choose a wide class of generic axisymmetric spacetimes that contain a black hole and admit a separable Hamilton-Jacobi equation of geodesics. Working in full generality, we find the equations of motion for the photon ring while obtaining the integration constants. To construct the Penrose limit, we obtain the Killing tensors of the spacetimes, and we further constrain ourselves to the subclass of spacetimes that admit Killing-Yano tensors, which we construct. Eventually, we construct the Penrose limit in terms of the initial metric spacetime elements for a large class of black holes. Our formalism and our final results are readily applicable to spacetimes that include a broad range of stationary black holes, such as Kerr, deformed Kerr, Kerr-Newman, exponentially suppressed mass-deformed Kerr, as well as various static black holes like Schwarzschild, Reissner–Nordström, Kiselev, and many others. We obtain the photon ring, the Lyapunov exponents, the orbital and precession frequencies, the quasinormal modes at high frequency limit,  and the wave equations in full generality with respect to the Penrose plane wave metric.

The paper is organized as follows. In section 2 we discuss the null geodesics in plane waves and their properties; in section 3 we discuss the separability of null geodesics and the photon rings; in section 4 the Penrose limit and the correspondence between the Lyapunov exponents, the orbital frequency and the quasinormal modes; in section 5 the photon rings as effective holographic horizons, potential bounds and the harmonic oscillator; in section 6 our conclusions. The paper is supplemented by an Appendix on the Unruh temperature of the Schwarzschild ring.

\section{Null Geodesics and the Harmonic Oscillator in Plane Waves}

In this section we briefly describe the plane wave of a spacetime. For an extensive review we refer the reader to \cite{blau} and for more details on the wave equation, we direct the reader to our subsection \ref{scalareq}. We consider a four-dimensional spacetime with  metric
\be
ds^2=g_{\mu\nu} dx^\m dx^\n~,
 \ee
and we assume that it admits  a covariantly constant Killing null vector field $V=V^\mu \partial_\mu$ such that $\nabla_\m V^\n=0$.
We may choose  coordinates adapted to $V$ so that $ V=\pp_v$ where $v$ is the coordinate along the curves of $V$. A general class of metrics admitting such a  parallel vector in coordinates $x^\m\rightarrow \{ u,v,x^m\}$, can be written as
\be \la{pp1}
ds^2=2 du dv+K_0(u,x^b)du^2+2 A_{a}(u,x^b)dx^a du+\d_{ij}dx^i dx^j~,
\ee
where we have used the fact that the existence of the vector $V$ we described, implies that $\pp_v g_\mn=0$ and that locally exists a function $u(x^\m)$ such that $V_\m=g_{v\m}=\pp_\m u$. In addition, we have constrained ourselves to the subclass of metrics with $g_{ij}=\delta_{ij}~$. In the metric  \eq{pp1}, the $u=$ constant wave front is flat and the full metric \eq{pp1} is called the plane-fronted wave with parallel propagation (pp-wave).

Limiting further the metric within the pp-wave subclass we may consider $A_\a=0$ and $K_0(u,x^b)\sim x^2$, to obtain the plane wave metric written in Brinkmann coordinates
\be \la{pw}
ds^2=2 du dv +A_{ij}(u) x^i x^j du^2+d\vec{x}^2~.
\ee
The plane wave metric depends on  the symmetric matrix $A(u)$ almost uniquely. The reason is that it is difficult to construct a coordinate transformation that maintains the form of the plane wave metric invariance, while respecting the choice of the gauge conditions, fact that gives a unique significance to the matrix $A$ in Brinkmann  coordinates.

The geodesics of the plane wave metric satisfy  equations of motion of a   non-relativistic oscillator with time depended frequency. In particular, null geodesics satisfy
\be\la{lag1}
-\dot{u} \dot{v}=\ff{1}{2} A_{ij} x^i x^j \dot{u}^2+\ff{1}{2}\dot{\vec{x}}^2~,
\ee
whereas the light cone momentum $p_v=\dot{u}$ is conserved,  giving rise to the light-cone gauge
\be
u=p_v \t~.
\ee
Similarly, the geodesic equations for the transverse coordinates $x^i$ are given by
\be \la{om0}
\ddx(\t)^i=A_{ij} p_v^2 x^j :=-\o^2_{ij}x^j~,
\ee
where the (time depended) frequency matrix $\omega_{ij}$ is identified as
\be \la{om1}
\o^2_{ij}=-A_{ij}(\t) p_v^2~.
\ee
One recognizes in Eq. (\ref{om0}) the equation for the non-relativistic harmonic oscillator.
As we can see from Eq. (\ref{om1}),  the frequency matrix is determined completely by the plane wave metric, and the  corresponding Hamiltonian can be written as
\be
H(\t)=\ff{1}{2}\prt{\dot{x}_i^2-p_v^2 A_{ij}(\t) x^i x^j}~.
\ee
One may further consider the geodesic deviation equation on families of geodesics, which also  described by  corresponding harmonic oscillator for the plane waves. Indeed, the separation $\d x^i$ between nearby geodesics is given by the solution of the equation
\be
\ff{D^2}{D\t^2} \d x^\m=R^\m_{\n\l\r}\dx^\n\dx^\l\d x^\r,
\ee
where $\ff{D}{D\t}=\dx^\m\nabla_\m$ is the covariant derivative along the curve $x^\m$.  In particular, the deviation of geodesics in the transverse directions satisfies   the equation
\be \la{om3}
\ff{d^2}{du^2}\d x^i=-p_v^{-2}R^i_{uju}\d x^j~,
\ee
where $R^i_{ubu}$ is the Riemann curvature tensor of a plane wave metric.
Since $R^i_{ubu}$ is the only non-vanishing component of the Riemann curvature tensor  given by
\be
R_{uiuj}=-A_{ij}~,
\ee
 the  geodesic deviation equation is written as
\be \la{om00}
\ff{d^2}{du^2}\d x^a=p_v^{-2}A_{ij}\d x^j~.
\ee
Clearly, the behavior of the geodesics in the transverse directions as well as their deviations depend on the signs of the $A_{ij}$. For example positive eigenvalues of $A$ (Lyapunov exponent) make the tidal forces repulsive while negative ones make it attractive.

For space-times that satisfy the vacuum Einstein equations, it can easily be verified that there is always one stable and one unstable direction for the transverse coordinates.  The only non-zero component  of the Ricci tensor   for a plane wave metric is
\be
R_{uu}=-{\rm Tr}\,  A~,
\ee
and  the vacuum Einstein equation then leads to a vanishing trace of the frequency matrix  $A_{ij}$, i.e.,
\be \la{tr}
{\rm Tr}\, A=0~.
\ee
Therefore, there is always a positive and a negative  eigenvalue of the frequency matrix.
This is a general rule for pp-wave backgrounds satisfying the vacuum Einstein equations  without the presence of null fluxes and matter. Moreover, if $A_{ij}(u)$ is constant, the Riemann curvature tensor is covariantly constant so that   the plane wave is locally symmetric.

\vskip.1in
\noindent

\section{Black Holes and Separability of Null Geodesics}

 In this section we will consider a wide class of black hole spacetimes, for which we will determine their Penrose limits and  we will investigate the stability of their null geodesics. The inverse of the instability timescale is the Lyapunov exponent. We will show that for the chosen class of black holes, the entire task of obtaining the Lyapunov exponent and associating it with the quasinormal modes boils down to the ability to find the Penrose limit of the black holes around the photon ring.

The Hamilton-Jacobi (H-J) equation of null geodesics is separable for the general stationary axisymmetric spacetime with  metric \cite{Papadopoulos:2018nvd} written in $(t,\th,\phi,r)$ coordinates of  the form
\be \label{metric1}
g_{\mn}(r,\theta) =
\left[\begin{array}{cccc}
-\ff{C_1 C_3}{C_4^2-C_3 C_5} & 0 & \ff{C_1 C_4}{C_4^2-C_3 C_5} & 0\\
0 & \ff{C_1}{B_2} & 0 & 0\\
\ff{C_1 C_4}{C_4^2-C_3 C_5} & 0 & -\ff{C_1 C_5}{C_4^2-C_3 C_5} & 0\\
0 & 0 & 0 & \ff{C_1}{A_2}
\end{array}\right]~ ,
\ee
where $C_i=A_i(r)+B_i(\th)$. Such spacetimes have at least two Killing vectors, one associated to time translation symmetry and one associated to axial symmetry. It follows that along these directions, we have two integrals  of motion for the geodesics, $E$ and $L$, respectively.

The Lagrangian for the geodesics in such a background reads
\be
2\cL= g_{tt} \dt^2+g_{rr} \dr^2+g_{\phi\phi} \dphi^2+g_{\th\th} \dth^2+2 g_{t\phi}\dt\dphi~.
\ee
The two generalized momenta
\be
p_t=g_{tt} \dt+g_{t\phi}\dphi=-E~,\qquad p_\phi=g_{\phi\phi}\dphi+g_{t\phi}\dt=L~,
\ee
are constant, since the metric is independent of $(t,\phi)$. The rest of the momenta are
\be
p_r= g_{rr} \dr~,\qquad p_\th= g_{\th\th} \dth~,
\ee
and can be found by solving the equations of motion. The H-J equation for null geodesics is
\be
g^{\mn} \pp_\m S \pp_\n S =0~,
\ee
with $p_\m=\pp_\m S$. Separability for motion in the geometry \eq{metric1} suggests the following ansatz
\be
S=-E t+ L \phi +S_r(r)+S_\th(\th)~,
\ee
which gives
\be
g^{rr}\pp_r S ^2+g^{tt}\pp_t S ^2+g^{\th\th}\pp_\th S ^2+g^{\phi\phi}\pp_\phi S ^2+2g^{t\phi}\pp_t S \pp_\phi S=0~,
\ee
and leads to two separable equations depending on $r$ and  $\th$
\bea \la{eqr}
&&A_2 S_r'^2+A_5 E^2+A_3 L^2-2A_4 E L=-\cK~,\\\la{eqth}
&&B_2 S_\th'^2+B_5 E^2+B_3 L^2-2B_4 E L=\cK~.
\eea
$\cK$ is an integration constant, and it is related to the usual Carter constant \cite{Carter:1968rr} up to a constant shift. Then the equations can be solved  for
\be\la{alleqs}
S'(r)=\ff{\sqrt{R(r)}}{A_2}~,\qquad S'(\th)= \ff{\sqrt{\Th\prt{\th}}}{\sqrt{B_2}}~,
\ee
with
\be \nn
R(r):=\sqrt{A_2} \sqrt{-\cK -A_5 E^2- A_3 L^2+2 A_4 E L}~,\quad \Th(\th):=\sqrt{\cK- B_5 E^2-B_3 L^2+ 2 B_4 E L}~.
\ee
The photon ring geodesics are located at $r=r_0$ and satisfy
\be \la{ring}
R(r_0)=0~\mbox{~and~}~ R'(r_0)=0~.
\ee
It is clear that we can introduce the constants
\be
b=-\ff{p_\phi}{p_t}=\ff{L}{E}~,\qquad  k_E=\ff{\cK}{E^2}~,
\ee
to reduce the total number of constants by a rescaling, where $b$ is the usual impact parameter defined for rotational orbits.  On the photon ring \eq{ring}, the constant $b$ satisfies
\be \la{eqb}
b=\prt{\ff{A_4'}{A_3'}\pm \ff{\sqrt{A_4'{}^2-A_3' A_5'}}{A_3'}}\Bigg|_{r_0}~,
\ee
while the constant $k_E$ reads
\be \la{eqck}
k_E=-A_5+2 \ff{A_4'}{A_3'}\prt{A_4-A_3 \ff{A_4'}{A_3'}}+\ff{A_3 A_5'}{A_3'}\pm 2 \ff{\sqrt{A_4'{}^2-A_3'A_5'}\prt{A_3 A_4'-A_3' A_4}}{A_3'(r)^2}\Bigg|_{r_0}~.
\ee
The equations \eq{eqb}, \eq{eqck} are effectively algebraic equations, readily applicable for any metric background of the form \eq{metric1} and together with   equations \eq{alleqs}, they determine the radial position of the photon ring and the parameters $(b, k_E)$.
Notice that, if we want to match the integration constant  with the usual Carter one for the Kerr black hole geodesics which is a special case of our analysis, a shift of $\cK\rightarrow \cK+\prt{b-a}^2$ is required.

\section{The Penrose Limit}

 To construct the Penrose limit for the class of the black hole spacetimes with metric (\ref{metric1}),  we need to find a tetrad of covectors that is parallel transported along the null geodesics of the photon ring:
 \be
 u^\m\nabla_\m e^{(i)}_\n=0~.
 \ee
The geodesic parallel transports its 4-velocity, and therefore this is the first leg of the tetrad. The second leg can be found by the contraction of the Killing-Yano tensor $Y$ with the 4-velocity: $Y_\mn u^\n$. To complete the tetrad, the two remaining legs are constructed to lie in the plane transverse to the two existing covectors. As a result, the non-triviality of the tetrad construction, boils down to the construction of the Killing-Yano tensor.

\subsection{Killing-Yano Tensors}

The second order  Killing-Yano tensors $Y_{ab}$ are antisymmetric tensors which satisfy the Killing-Yano equation
\be \la{KY_eq}
\nabla_{c}Y_{ab}+\nabla_{a}Y_{cb}=0~,
\ee
from which it  follows that the $Y_\mn u^\n$ is parallel transported along a geodesic. In addition, it  satisfies
\be \la{KYK}
K_{ab}=Y_a^{~c} Y_b^{~c},
\ee
where the tensor $K_{ab}$ is a Killing tensor
\be
\nabla_{(a}K_{bc)}=0~.
\ee
In order to solve the Killing-Yano equation, we make a choice for  the  functions $A_i(r)$ and $B_i(\th)$ in the metric \eq{metric1} as follows
\bea \nn
&&A_1=\Xi(r)^2~,\quad A_2=\D(r)~,\quad A_3(r)= -\ff{a^2 \Phi(r)}{\D(r)}~,\\ &&A_4(r)=-\ff{a \Phi(r) \prt{\Xi^2(r) +a^2}}{\D(r)}~,\quad A_5(r)=\ff{A_4(r)^2}{A_3(r)}~.\la{ansatzais}
\eea
The $\theta$ dependent functions in the metric elements are chosen as in the Kerr geometry in order to include the Kerr spacetime as a special case of our metrics, and therefore are given by
\be \la{ansatzbis}
B_1(\th)=a^2 c_\th^2~,\quad B_2(\th)=1~,\quad B_3(\th)=s_\th^{-2}~,\quad B_4(\th)=a~,\quad B_5(\th)=\ff{B_4(\th)^2}{B_3(\th)}~,
\ee
where the $s_\th$ and $c_\th$ denote respectively the sine and cosine of the angle $\th$. A similar ansatz has been considered in \cite{Baines:2023dhq}.
In the standard orthonormal basis,
\bea
&&e^1=\sqrt{\ff{\Xi(r)^2+a^2c_\th^2}{\D(r)}} dr~,\quad e^2=\sqrt{\Xi(r)^2+a^2c_\th^2} d\th~,\\ \nn
&&
e^0=\ff{\sqrt{\D(r)}}{\sqrt{\Phi(r)\prt{\Xi(r)^2+a^2c_\th^2}}}\prt{dt-a s_\th^2d\phi}~,\quad e^3=\ff{s_\th}{\sqrt{\Xi(r)^2+a^2c_\th^2}}\prt{-a dt+ \prt{\Xi(r)^2+a^2}d\phi}~,
\eea
the Killing tensor is found to be  the diagonal
\be \la{killing1}
K^{AB} ={\rm diag}\prt{-a^2 c_\th^2, a^2 c_\th^2,-\Xi(r)^2,-\Xi(r)^2}~.
\ee
Then,  Eq. \eq{KYK} implies that $Y$ is of  the form
\be
Y=s_1 a c_\th e_0\wedge e_1+s_2 \Xi e_2\wedge e_3~, \qquad s_i=\pm 1
\ee
and  by switching back to coordinate basis, we find the solution for covariant components that satisfies  Eq. \eq{KY_eq}  as
\be \la{KYmatrix}
Y(r,\theta) =
\left[\begin{array}{cccc}
0 & s_2 a \Xi s_\th  & 0 & \ff{s_1 a c_\th}{\sqrt{\Phi}} \\
-s_2 a \Xi s_\th  & 0 & s_2 \Xi(a^2 +\Xi^2)s_\th & 0\\
0& -s_2 \Xi (a^2 +\Xi^2)s_\th& 0 & -\ff{s_1 a^2 c_\th s_\th^2}{\sqrt{\Phi}}\\
-\ff{s_1 a c_\th}{\sqrt{\Phi}} & 0 & \ff{s_1 a^2 c_\th s^2_\th}{\sqrt{\Phi}}& 0\\
\end{array}\right] ,
\ee
with
\be \la{cond0}
\Xi(r)=-\ff{s_2}{s_1}\int\ff{1}{\sqrt{\Phi(r)}}dr+c_1~,
\ee
and arbitrary $\D$. The requirement of the existence of the Killing-Yano tensor reduces the number of independent functions describing the black hole spacetime to two: $\Phi$ and $\D$.
Still, the class of the black holes under study is vast. For our purposes, let us choose the solution
\be \la{cond}
\Phi(r)=1~, \qquad \Xi(r)=\pm r~,
\ee
for which
\be \la{KYmatrix2}
Y(r,\theta) =
s_1\left[\begin{array}{cccc}
0 & -a r \sin\th & 0 & a c_\th \\
a r s_\th & 0 &- r(a^2 +r^2)s_\th & 0\\
0& r (a^2+r^2) s_\th & 0 & -a^2 c_\th s_\th^2\\
-a c_\th & 0 & a^2 c_\th s^2_\th& 0\\
\end{array}\right] ~,
\ee
while $\D(r)$ remains always arbitrary.  This class of metrics includes the stationary black holes: Kerr, deformed Kerr, Kerr-Newman, exponentially suppressed mass deformed Kerr and other type of deformations, as well as the static black holes: Schwarzschild, Reissner–Nordström, Kiselev and many other. The explicit form of the corresponding metric can be easily found by substituting the above solutions \eq{cond} to the metric \eq{metric1}. For completeness we present its form in the Appendix \ref{appA0}. Moreover, note that a subclass of the neutral black holes we consider can be seen as special cases of the Kerr-NUT-de Sitter spacetimes, where their Killing-Yano tensors have been discussed in  \cite{Houri:2007xz}.  Next, we will find the Penrose plane wave limit for the class of metrics satisfying \eq{cond}.

\subsection{The Penrose limit}

Having the Killing-Yano tensor, we are able to construct the full tetrad that is parallel transported along the null geodesics and comprise the photon ring. The resulted metric is given by
\be \la{pwmetric}
ds^2=2 du dv +A_{ij}x^i x^j du^2+dx_1^2+dx_2^2~,
\ee
where
\be\la{Aplane}
A_{ij}=-R_{\mn\a\b}u^\m  e^{(i) \n} u^\a  e^{(j)\b}\big|_{\rm null~geodesic}~.
\ee
The construction of the parallel frames for the null geodesics gives
\be \la{s1}
u_\m=\pp_\m S~, \quad u_\m u^\m=n_\m n^\m=0~,\quad e_\m^{(i)} e^{\m (j)}= \d^{ij}~,
\ee
where \cite{Kubiznak:2008zs}
\bea\la{tetrad1}
&& e^{(1)}{}^{\mu} = \frac{1}{C}\prt{ u^{\a} h_{\a}{}^{\mu} - u (u^{\a}\xi_{\a})u^{\mu}}~, \qquad 	e^{(2)}{}^{\mu} = \frac{1}{K}\prt{u^{\a} Y_{\a}{}^{\mu} } ~,\\\nn
&&n^{\mu} =
\frac{1}{C}  e^{(1)}{}^{\a } h_{\a }{}^{\mu} + \frac{1}{2 C^4} \left( C_{\beta}{}^{\gamma}C_{\g  \delta} u^{\b } u^{\d } + u^2 (\xi_{\a } u^{\a })^2 C^2 \right) u^{\mu}~.
\eea
We have defined above  the conformal Killing-Yano tensor $h_{\mu\nu}$ as  the Hodge dual of the Killing-Yano tensor $Y$
\begin{align}
h=\star Y=h_{\mu\nu}d x^\mu x^\nu&=-s_1 a\prt{a^2+r^2} c_\th s_\th d\th \wedge d\phi- s_1 r dt\wedge dr \nonumber \\
&- s_1 a^2 c_\th s_\th dt \wedge d\th+ s_1 a r s_\th^2 d\phi
\wedge dr~.
\end{align}
 In addition, we have defined
\be
\xi^\a=\frac{1}{3} \nabla_\b h^{\b\a}=\prt{-s_1,0,0,0}, \qquad
C_{\a \b} = h_{\alpha \gamma} h_{\b}{}^{\g},
\ee
and
 \be
 K^2 = K_{\a \b} u^{\a} u^{\b} ~ , \qquad  C^2 = C_{\a \b} u^{\a} u^{\b} ~.
 \ee
Since  our final results depend on $s_i^2$, without loss of generality, we may assume that  $s_1=-s_2=1$.
The contraction of the conformal Killing tensor  reads
\be
C^2=\frac{a^2 \D(r_0)\left((a s_{2\th}-2 b \cot (\th))^2+4 c_\th^2 S'(\th)^2\right)+4 r^2\prt{ \left(a (a-b)+r_0^2\right)^2-\D(r_0)^2}}{2 \D(r_0) \left(a^2 c_{2 \th}+a^2+2 r_0^2\right)}~,
\ee
which takes the very simple form for equatorial geodesics
\be
C^2=-\ff{\prt{ a\prt{a- b}+r_0^2}^2}{\D(r_0)}-\D(r_0)~.
\ee
The Killing tensor can be found in coordinate basis by using equation \eq{killing1}. Its contraction with the null four velocity turns out to simplify for the equatorial geodesics as
\be
K^2=\prt{a-b}^2~.
\ee
Similarly, the contraction of the conformal Killing tensor along the null velocity is
\be \nn
C_{u_2}:=C_{\beta}{}^{\gamma}C_{\g  \delta} u^{\b } u^{\d }=\frac{a^4 \D(r_0) c_\th^2\left((a s_{2\th}-2 b \cot (\th))^2+4 c_\th^2S'(\th)^2\right)-4 r_0^4 \left(a^2-a b+r_0^2\right)^2}{2 \D(r_0) \left(a^2 c_{2\th}+ a^2+2 r_0^2\right)}
\ee
which at the equatorial plane is written as
\be
C_{u_2}=-\frac{r_0^2 \left(a  (a-b)+r_0^2\right)^2}{\D(r_0)}~.
\ee
We point out that all the above quantities are computed at the photon ring $r=r_0$.

At this stage all the information to construct the parallel propagated tetrad \eq{tetrad1}  is known. We refrain from providing the long expressions for out of equatorial plane geodesics and present the equatorial case, using the information of the null geodesics given by equations \eq{alleqs}, \eq{eqb} and \eq{eqck}.

The impact parameters can be found by applying the equations \eq{eqb} and \eq{eqck} to be
\be \la{kshifted}
 a b=a^2+r^2-4 r\ff{\D(r)}{ \D'(r)}~,\qquad a^2 k_E=16 a^2 r^2\ff{\D(r)}{\D'(r)^2}-\ff{r^2 \prt{r \D'(r)-4 \D(r)}^2}{ \D'(r)^2}~,
\ee
where we have redefined $k_E$ by the constant shift $\prt{b-a}^2$: $k_E\rightarrow k_E +\prt{b-a}^2$.
The constant $k_E$ enters in the $\Theta(\th)$ equation of motion \eq{alleqs}, where for the equatorial geodesics we have
\be
k_E=0~.
\ee
Then, Eq. (\ref{kshifted}) is written as
\begin{eqnarray}
&&16 a^2 r^2\ff{\D(r)}{\D'(r)^2}-\ff{r^2 \prt{r \D'(r)-4 \D(r)}^2}{ \D'(r)^2}=0, \la{kshifted2}\\
&& a b=a^2+r^2-4 r\ff{\D(r)}{ \D'(r)}~.\la{kshifted1}
\end{eqnarray}
Eq. (\ref{kshifted2}) determines the location of the photon ring $r_0$, whereas  Eq. (\ref{kshifted1}) specifies the value of $b$ at the photon ring.
It is worthy to present the even more compact expressions for the static black holes as a special case
\be \la{kshiftedazero}
b=\ff{r^2}{\sqrt{\D(r)}}~,\qquad \D'(r)=\ff{4 \D(r)}{r}~,
\ee
where again the solution of the second algebraic equation determines the position $r_0$ of the photon ring and the first one the impact parameter $b$.

Finally, using the Eqs. \eq{kshifted} we can simplify the expression for $A_{ij}$, which now reads
\bea \nn
&&A_{11}=
\ff{4}{r_0^2}-8\ff{ \D \prt{ r_0} \prt{r_0 \D''(r_0)-\D'(r_0)}}{r_0^3 \D'(r_0){}^2}~,\\\la{Ais}
&&A_{22}=
-\ff{2}{r_0^2}+8\ff{ \D(r_0) \prt{\D'(r_0)-2 r_0}}{r_0^3 \D'(r_0){}^2}~, \\\nn
&&A_{21}=A_{12}=0~.
\eea
For static black holes, the plane wave metric turns out to be much simpler
\bea \la{Aisqzero}
A_{11}=\ff{6}{r_0^2}-\ff{\D''(r_0)}{2 \D(r_0)}~,\qquad A_{22}=-\ff{1}{\D(r_0)}~,\qquad
A_{21}=A_{12}=0~.
\eea
Here we note that the plane wave for equatorial orbits is always diagonal for the class of separable metrics.
Moreover, the diagonal elements $A_{ii}$ are not of equal norm in the generic case, since we have not yet imposed any constraint on the original background. We will confirm below that a solution to the Einstein equations without the presence of null matter or fluxes leads always to $A_{11}=-A_{22}$.  To rephrase the above statement, for the vacuum Einstein solutions, the equation $A_{11}+A_{22}=0$, using for example \eq{Aisqzero}, is an algebraic equation that provides the position of the photon ring $r_0$.

\subsection{Lyapunov Exponents and Orbital Frequencies}

 In order to obtain the Lyapunov exponent and the frequencies we are interested in, we need to find the relation between the Brinkmann coordinates and the initial coordinate system of the full spacetime.  The transverse directions in the Penrose limit are along $r$ and $\th$. The solution to the geodesic in the initial coordinate system of the black hole is
\be \la{geo}
r(u)=r_0~,\qquad \th(u)=\ff{\pi}{2}~,\qquad t(u)= p_v^{-1} u~,\qquad \phi(u)= \tilde{\o}_\phi u= \tilde{\o}_\phi p_v t~,
\ee
where $u$ is an affine parameter.  The Lyapunov exponent can be read by the null geodesic equations of motion, and the orbital frequency from \eq{geo}. Both are given by
\be
\l^2= A_{11} p_v^2~,\qquad \o_{orb}=\tilde{\o}_\phi p_v~.
\ee
The expressions for $p_v$ and $\tilde{\o}_\phi$ can be found in full generality. For stationary black holes we have
\bea \la{lyapan0}
&&\tilde{\o}_\phi=\ff{1}{\sqrt{\D(r_0)}}~, \quad  \l^2=\ff{64\D(r_0)^2\prt{r \D'(r_0)^2+2\D(r_0)\prt{\D'(r_0)-r\D''(r_0)}}}{r^3\prt{4\D(r_0)\prt{4 r-\D'(r_0)}+r\D'(r_0)^2}^2}~,\\\nn
&&p_v=\ff{4 \D(r_0)\D'(r_0)}{4 \D(r_0)\prt{4 r_0-\D'(r_0)}+r_0 \D'(r_0)^2}~, \quad \o_{orb}^2=\ff{16 \D(r_0) \D'(r_0)^2}{\prt{4\D(r_0)\prt{4 r_0-\D'(r_0)}+r\D'(r_0)^2}^2}~,
\eea
where we have used the photon ring equation to trade the rotation parameter $a$ with the photon ring radius $r_0$. For static black holes, the Lyapunov exponent takes the much simpler form
\be \la{lyapa0}
p_v=\ff{\D(r_0)}{r_0^2}~,\quad \tilde{\o}_\phi=\ff{1}{\sqrt{\D(r_0)}}~,\quad \l^2=\ff{\D(r_0)\prt{12 \D(r_0) -r_0^2 \D''(r_0)}}{2 r_0^6}~,\quad \omega_{orb}^2=\ff{\D(r_0)}{r_0^4}~.
\ee
Moreover, there is an additional frequency $\o_{prec}^2=\l^2$ describing perturbations in the transverse stable $\th$ direction which cause a precession around the original orbit. The equality of the precession frequency with the Lyapunov exponent comes from the transverse fluctuations of motion.

\subsection{Massless Scalar Equation in Penrose plane wave limit}\label{scalareq}

The massless scalar wave equation is
\be
\ff{1}{\sqrt{-g}}\pp_\m \prt{\sqrt{-g} g^{\mu\nu} \pp_\n \Psi}=0~.
\ee
It is known that it is separated in the original metric when $[K,R]^\m_{~\n}$ is a covariantly constant tensor \cite{2002JMP43,Giorgi:2021skz}
\be\la{kgs}
\nabla_\m [K,R]^\m_{~\n}=\nabla_\m\prt{K^\m_{~l} R^l_{~\n}-R^\m_{~l} K^l_{~\n}}=0~.
\ee
It can be shown  that for the spacetimes we consider,  the existence of the Killing-Yano tensor and Eq. \eq{cond}, the  equation \eq{kgs} is satisfied, and it leads to the separability of the wave equation.
Therefore, we can write the solution  for a diagonal $A$ as
\be
\Psi=e^{i \prt{p_v v+p_u u}}\psi_1(x_1) \psi_2 (x_2)~,
\ee
which gives rise to the following  two independent equations for $\psi_1(x_1)$ and $\psi_2(x_2)$
\bea \la{wave1}
&&\ff{1}{2 p_v^2} \psi_1''(x_1)+\ff{1}{2} A_{11}x_1^2\psi_1(x_1)=\prt{\ff{p_u}{2 p_v}+c_0}\psi_1(x_1)~,\\
&&\ff{1}{2 p_v^2} \psi_2''(x_2)+\ff{1}{2} A_{22}x_2^2\psi_2(x_2)=\prt{\ff{p_u}{2 p_v}-c_0}\psi_2(x_2)~,
\eea
where we have denoted as $c_0$ the separation constant. The above equations describe the quantum harmonic oscillator. Here our unstable direction is the $x_1\sim \d r$ and the stable the $x_2\sim \d \th$. For the quasinormal modes and in order to solve the above equations, we impose a decaying and an outgoing boundary condition for the stable and the unstable direction, respectively.   In general we can think of the frequencies as having the form
\be
\o_{L n_1 n_2}\sim L \o_{orb}+\prt{\ff{1}{2} +n_1}\o_{prec}-i\l\prt{\ff{1}{2}+n_2}~,
\ee
where $\o_{orb}$ is the frequency of the stable orbits, $\l$ is the imaginary part of the frequency and equal to the Lyapunov timescale for unstable orbits.

In the original metric, in the large frequency limit the wave equation can be solved such that the wavefronts of the plane waves propagate along the nulls geodesics. The quasinormal modes' exponential decay is associated with the exponential deviation of the null geodesics. On the Penrose limit we are already parallel transported to the null geodesics. Moreover, the null geodesic deviation is identical with the ones of the original metric. Therefore, the wave solutions of the equation \eq{wave1} correspond to the null geodesics. In other words, one may read the Lyapunov coefficients of the null geodesics and the quasinormal modes directly from the plane wave Hamiltonian when computed on the photon sphere.

\subsection{Surface Gravity}

The surface gravity measures the gravitation force on a static observer at the horizon surface as seen by an observer at infinity. It can be thought that for static spherically symmetric black holes as being $\k\sim V \a~,$ where  $V$ is the  redshift factor and $\a$ is the acceleration. In several geodesic instabilities of different types of geodesics, the surface gravity plays a central role. In fact, it has been conjectured in  \cite{Hashimoto:2016dfz}, a possible classical bound of the Lyapunov exponent for particle geodesics related to the surface gravity, which has been motivated by an associated bound of the Lyapunov exponent in the   Out-of-Time Order Correlator (OTOC) of quantum chaos \cite{Maldacena:2015waa}. However, this bound can be violated,  for generic black hole spacetimes, even when all the geodesic and energy conditions are imposed \cite{Giataganas:2021ghs}.  This has been further confirmed for certain black holes, as for example \cite{Lei:2023jqv,Gwak:2022xje,Jeong:2023hom}. Furthermore, it has been  observed  \cite{Cardoso:2008bp}, that  the Lyapunov exponent of the unstable null geodesics of near-extremal Schwarzschild-de Sitter geodesics, and therefore the imaginary part of the quasinormal frequencies, are related to the surface gravity. These developments naturally motivate the idea that a relation between the Lyapunov exponent of the geodesics and the surface gravity exists.  Here we will show that we can express our results in terms of the surface gravity in the black hole spacetimes   when certain limits are considered.

The surface gravity is computed by the contraction of the covariant derivatives of the Killing vectors
\be \la{killing}
K^\m=\prt{1,0, \ff{a}{\Xi(r_h)^2+\a^2},0} ~,
\ee
with vanishing norms at the Killing horizon. The two non-zero components of the vector are associated with the axial symmetry and the time translation of our black hole spacetime. The surface gravity then reads
\be
\k^2=-\ff{1}{2}\prt{\nabla_\m K_\n}\prt{\nabla^\m K^\n}\bigg|_{r_h}~,
\ee
and substitution of \eq{killing} leads to
\be\la{surf1}
\k=\ff{\D'(r_h)}{2  \prt{r_h^2+a^2}}~.
\ee
If we want to restrict to spherically symmetric static black holes, the above expression turns out to be
\be
\k=-\ff{g_{tt}'}{2\sqrt{-g_{tt} g_{rr}}}\bigg|_{r_h}~.
\ee
For a  Schwarzschild black hole in particular, (\ref{surf1}) simplifies to $\k_{Schw}=\prt{4m}^{-1}=2 \pi T$.

\subsection{Application of the Formalism to Specific Black holes}

For static and stationary black holes the rotation parameter $a$ related to the angular momentum is null, and  the photon ring equations are simplified to \eq{kshiftedazero}.
The solution of the second of these algebraic equations determines the radius of the photon ring, whereas  the first one specifies the value of the impact parameter. The Penrose limit on the photon ring is given by the metric (\ref{pwmetric}), where the matrix $A_{ij}$ is given by Eq.  \eq{Aisqzero}.

For stationary black holes  the photon ring equations are given by,  Eq.(\ref{kshifted2}) which determines the location of the photon ring $r_0$, whereas  Eq. (\ref{kshifted1})  the impact parameter. The matrix elements $A_{ij}$ of the Penrose limit on the photon ring are given by  Eq.  \eq{Ais}.

In this section we apply our formalism to a wide range of static or stationary  black holes.

\subsubsection{Schwarzschild Black Holes}
Let us start with the static Schwarzschild black hole. The only data we have to provide is
\be
\D(r)=r^2-2 m r~,\qquad a=0~,\qquad r_h=2 m~.
\ee
By solving  the equations \eq{kshiftedazero}, we find that the location of the photon ring and the impact parameter are given by
\be
r_0=3 m~,\qquad b^2=27 m^2~.
\ee
By applying the equations \eq{Aisqzero} we obtain the metric elements of the Penrose limit plane wave in Brinkmann coordinates
\be
A_{11}=-A_{22}=\ff{1}{3 m^2}~,\qquad A_{12}=A_{21}=0~.
\ee
The direction $x_2\sim \th$ is stable: $A_{22}\le 0$, while the direction $x_1\sim r$ is unstable: $A_{11}\ge 0$. The matrix $A$ can expressed with respect to the surface gravity $\k_{Schw}=\ff{1}{4 m}$ as
\be
A_{11}=\ff{16}{3}\k_{Schw}^2~.
\ee
The Lyapunov exponent and the orbital frequency are \eq{lyapa0}
\be \la{lschw}
\l^2=\o_{orb}^2=\ff{1}{27 m^2}=\ff{16}{27} \k_{Schw}^2\le \k_{Schw}^2 ~.
\ee
In the neutral static black hole where only one dimensionful parameter exist, we can trivially trade the surface gravity with the entropy of the black hole, $S=Area/4$ to rewrite the above expression as
\be
\l^2=\ff{4 \pi}{27 S}\le\ff{\pi}{4 S}~,
\ee
which allows us to express the Lyapunov exponent and the frequencies with respect to the entropy. The inequality has been motivated by the fact that $\k_{Schw}^2= \pi/(4S)$ and its importance will become more clear in black holes with more parameters in the next section \ref{section:holo}.

\subsubsection{Reissner–Nordström Black Hole}
For the Reissner–Nordström Black hole we have
\be
\D(r)=r^2-2 m r+q^2~,\qquad a=0~.
\ee
The horizons are located at
\be
r_{h\pm}=m\pm \sqrt{m^2-q^2}
\ee
and the surface gravity is \eq{surf1}
\be
\k_{RN}=\ff{\sqrt{m^2-q^2}}{\prt{m+\sqrt{m^2-q^2}}^2}~,
\ee
with $m\ge |q|$.   Applying the equations \eq{kshiftedazero}, we obtain the radii for the photon ring
\be
r_{0\pm}=\ff{1}{2} \prt{3 m \pm\sqrt{9 m^2-8 q^2}}~,\qquad b^2=\ff{2 r_0^3}{r_0-m}~,
\ee
where $r_{0+}>r_{h+}$ is the radius of the outer ring, while the inner solution for the circular ring is inside the horizon $r_{h-}<r_{0-}<r_{h+}$.
From the equations \eq{Ais}, we find that the matrix $A_{ij}$ for the Penrose limit plane wave is
\begin{align}
A_{11\pm}&=\frac{2}{3 m^2-q^2\pm m\dfrac{9 m^2-7  q^2}{\sqrt{9 m^2-8 q^2}}}, \nonumber \\
A_{22\pm}&=\frac{2}{2 q^2- 3 m^2 \mp\sqrt{9 m^2-8 q^2}}~,~\qquad A_{12}=A_{21}=0~.
\end{align}
 $A_{ii\pm}$ corresponds to $r_{0\pm}$ and clearly, $A_{11\pm}\neq -A_{22\pm}$, unless $q=0$ as expected. For completeness we have presented the $A_{ij}$ for both photon rings, but let us  focus in the following on the outer ring.

For massive black holes the radius of the photon ring is $r_{0+}\simeq3 m\prt{1-\ff{2 q^2}{9 m^2}}$ and the Penrose limit matrix $A$ is given by
\be
A_{11}\simeq \ff{16}{3}\k_{RN}^2\prt{1+\ff{16}{3} q^2 \k_{RN}^2}~,\quad A_{22}\simeq -\ff{16}{3}\k_{RN}^2\prt{1+\ff{80}{9} q^2 \k_{RN}^2}~.
\ee
Note that the $q$-dependent terms are responsible for the different values of $A_{ij}$ in the charged black hole.
The Lyapunov exponent and the orbital frequency \eq{lyapa0} are found to be
\be
\l^2\simeq\ff{16}{27} \k_{RN}^2\prt{1+\ff{16}{9} q^2 \k_{RN}^2}~,\qquad \o_{orb}^2=\ff{16}{27} \k_{RN}^2\prt{1+\ff{16}{3} q^2 \k_{RN}^2}~,
\ee
where $\l\neq \o_{orb}$ in contrast to the uncharged black hole, since the $q-$dependent terms differ.

For the near extremal Reissner–Nordstr\"om black hole, $m\simeq |q|$ and the horizons $r_{h\pm}\simeq 2 m$ tend to become degenerate. In this case, we have for  the outer photon ring
\be
A_{11}\simeq\ff{1}{2 q^2}-\ff{\k_{RN}^2}{2}~,\qquad A_{22}\simeq -\ff{1}{q^2}+4\k_{RN}^2~,
\ee
so that the Lyapunov exponent and the orbital frequency are \eq{lyapa0}
\be
\l^2\simeq\ff{1}{32}\prt{\ff{1}{q^2} +\k_{RN}^2}~,\qquad \o_{orb}^2\simeq\ff{1}{8}\prt{ \ff{1}{2q^2}  -\k_{RN}^2}~,
\ee
with $\k_{RN}\ll1$. It follows that the nearly stable circular geodesics of the extremal charged black hole are at $r_{0+}=2 m$ and the impact parameter is  $b^2=16 m^2$. Notice that the photon ring is closer to the horizon of the black hole in the charged case compared to the uncharged one. The non-zero elements of the matrix $A$  in  the Penrose limit for the near extremal Reissner–Nordström black hole turn out then to be
\be
A_{11}=\ff{1}{2 m^2}~,\qquad A_{22}=-\ff{1}{m^2}~.
\ee
Therefore, the Lyapunov exponent and the orbital frequency are \eq{lyapa0}
\be
\l^2=\ff{ 1}{32 m^2} ~,\qquad \o_{orb}^2=\ff{1}{16 m^2}  ~,
\ee
which are both larger in the extremal limit than the surface gravity $\k_{RN}=0$, while for the inner ring in the extremal case $\l_{-}\rightarrow0$.

\subsubsection{Kerr Black Hole}

The Kerr black hole has
\be
\D(r)=r^2-2 m r+a^2,
\ee
and the horizons  are located at the radii
\be
r_{h\pm}= m \pm \sqrt{m^2-a^2},
\ee
and the surface gravity and temperature are \eq{surf1}
\be
\k_K=\ff{\sqrt{m^2-a^2}}{2 m\prt{m + \sqrt{m^2-a^2}}}~,\qquad T=\ff{\k_K}{2\pi}~.
\ee
Application of the generic analysis leads to the compact form of the Penrose limit matrix $A$
\be
A_{11}=-A_{22}=\ff{3}{r_0^2}~,
\ee
where $r_0$ are the two photon ring radii. By applying Eq. \eq{ring}, we find that
\be \la{radiikerr}
r_{0+}=2 m\prt{1 + \cos\prtt{\ff{2}{3}\arccos\prt{\ff{a}{m}}}}~,\quad r_{0-}=2m\prt{1+\cos\prtt{\ff{4 \pi}{3}+\ff{2}{3}\arccos\prt{\ff{a}{m}}}}~,
\ee
with $r_{0+}\ge r_{0-}$, corresponding to the prograde $(r_{0-})$ photon orbits which are at a smaller radius than the retrograde $(r_{0+})$ orbits, and spinning at opposite direction due to the Lense-Thirring effect. The direction or rotation for each photon ring becomes evident by computing the sign of $p_\phi$ or equivalently the parameter $b$ for the two photon rings. At the limit of zero angular momentum, we can check that we obtain smoothly $r_{0+}=r_{0-}=3m$. Note that the positions of the orbits are bounded by $m\le r_{0-}\le 3m\le  r_{0+}\le 4m$.
Furthermore, the Lyapunov exponent and the orbital frequency read
\be
\l^2=\ff{3\prt{m-r_0}^2}{r_0^2(3m+r_0)^2}~,\qquad \o_{orb}^2=\ff{4 m}{r_0\prt{r_0+3m}^2}~.
\ee
We find that the following inequality is satisfied for the external photon ring
\be \la{ker1a}
\o_{orb}^2\le \l^2,
\ee
which is saturated  for the static case $r_0=3m$.
We propose an overall bound for the Lyapunov exponent as
\be
\o_{orb}^2\le \l^2\le \ff{16}{27}\k_{+}^2~,\qquad \k_{+}:=\ff{r_{h+} +r_{h-}}{2\prt{r_{h+}^2+a^2}}
\ee
which approaches saturation only for the static black hole $a=0$.
\footnote{We point out that the inner photon ring satisfies the inverted bound inequality  $\o_{orb}^2\ge \l^2$. Therefore the inequality of the orbital frequency with the photon ring Lyapunov exponent, is a strong characteristic of each photon ring, and has consequences on the magnitudes of the real and imaginary parts of the quasinormal modes.}

In the limit of large mass, the Penrose limit is
\be
A_{11 \pm}\simeq \ff{1}{3 m^2}\prt{1\mp\ff{4a}{3\sqrt{3}m}}\simeq
\ff{16}{3}\k_{K}^2\prt{1\mp\ff{16 a \k_{K}}{3 \sqrt{3}}}~,
\ee
where for the outer retrograde photon ring, $A_{11+}$ is smaller than $A_{11-}$ of the inner ring.
For the outer photon ring, the Lyapunov exponent and the orbital frequency read now
\be \la{kerrl}
\l^2\simeq\ff{16}{27}\k_{K}^2\prt{1+\ff{152 a^2 \k_{K}^2}{27}}~,\qquad \o_{orb}^2\simeq\ff{16}{27}\k_{K}^2\prt{1-\ff{16 a \k_{K}}{3 \sqrt{3}}}~,
\ee
where we see that the first subleading term for $\l^2$ is positive, while for $\o_{orb}^2$ is negative. For rotating massive black holes, the Lyapunov exponent increases compared to the static ones, when expressed in terms of the surface gravity.

In the near extremal limit for the Kerr black holes we get
\be
A_{11+}\simeq\ff{3}{16 m^2}+\ff{\k_{K}^2}{6}~,\qquad A_{11-}\simeq\ff{3}{m^2}-\ff{8 \sqrt{3}\k_{K}}{m}~.
\ee
Let us focus on the extremal case, where  we have a maximal separation of the photon rings and a single horizon $r_h$
\be
r_{0-}=r_h=m~,\qquad  r_{0+}=4 m~.
\ee
The Penrose metric elements on the two photon rings maximize their distance apart as well
\be
A_{11+}=16 A_{11-}=\ff{3}{16 m^2}~,
\ee
while always the values of the inner photon ring are lower than the outer ring. The Lyapunov coefficient and orbital frequencies read
\be
\l_+^2=\ff{27}{784 m^2}~,\qquad \o_+^2=\ff{1}{49 m^2}~,\qquad \l_-^2=0~,\qquad \o_-^2= \ff{1}{4m^2}~,
\ee
with $\k_K\rightarrow 0$. As in the static charged black hole, the outer ring in the extremal limit has a Lyapunov exponent that is larger than the surface gravity, while in the inner ring it is trivially equal to it.

\subsubsection{Kerr-Newman Black Hole}

For the Kerr-Newman black hole we have
\be
\D(r)=r^2-2 m r+a^2+q^2,
\ee
and  the horizons are located at
\be
r_{h\pm}=m\pm\sqrt{m^2-a^2-q^2}~.
\ee
The surface gravity \eq{surf1} in this case is
\be
\k_{KN}=\ff{\sqrt{m^2-a^2-q^2}}{a^2+ (m- \sqrt{m^2-a^2-q^2})^2}~.
\ee
We solve the photon ring equations for $(a,b)$, and when we substitute to \eq{Ais}, we get the  metric elements, i.e. the matrix $A_{ij}$,  in  the Penrose limit  in a compact form
\be
A_{11}=\frac{4 q^2-3 m r_0}{r_0^2 \left(q^2-m r_0\right)}~,\qquad
A_{22}=\frac{-2 q^2+3 m r_0}{r_0^2 \left(q^2-m r_0\right)}~,\qquad A_{12}=A_{21}=0~.
\ee
Notice that in general $A_{11}\neq A_{22}$ unless $q=0$, where we recover the Penrose plane wave limit of the Kerr black hole. The Lyapunov exponents and the orbital frequency read
\be
\l^2=\frac{(m-r_0)^2 \left(4 q^2-3 m r_0\right)}{\left(q^2-m r_0\right) \left(r_0 (3 m+r_0)-2 q^2\right)^2}~,\qquad \o_{orb}^2=\frac{4 m r_0-4 q^2}{\left(r_0 (3 m+r_0)-2 q^2\right)^2}~.
\ee
For massive black holes we obtain for the outer ring
\bea \nn
&&A_{11}  \simeq\frac{16 \k_{KN}^2}{3}\prt{1-\frac{16 a \k_{KN}}{3 \sqrt{3}}+\frac{8}{27} \k_{KN}^2 \left(59 a^2+18 q^2\right)}~,\\
&&A_{22} \simeq-\frac{16 \k_{KN}^2}{3}\prt{1-\frac{16 a \k_{KN}}{3 \sqrt{3}}+\frac{8}{27} \k_{KN}^2 \left(59 a^2+30 q^2\right)}~,
\eea
where  $A_{11}$ and $-A_{22}$ differ by  $q-$dependent terms., which are responsible for the charged correction compared to the Kerr black hole data so that $\l$  and $\o_{orb}$ read
\begin{eqnarray}\la{knl}
\l^2&\simeq& \ff{16}{27}\k_{KN}^2\prt{1+\ff{8}{27}\prt{19\a^2+6 q^2}\k_{KN}^2}~,~  \\
\o_{orb}^2&\simeq& \ff{16}{27}\k_{KN}^2\prt{1-\ff{16}{3\sqrt{3}}\a\k_{KN}+\ff{8}{9}\prt{19 \a^2+6 q^2}\k_{KN}^2}. \nn
\end{eqnarray}
Note that, similarly to the effect of angular momentum, the charge $q$ increases further the Lyapunov exponent and the orbital frequency.

\section{Photon Ring as a Holographic Horizon and Bounds} \la{section:holo}

The plane wave metric and the Lyapunov exponent can be expressed through the surface gravity of the photon ring of the black hole. For the Schwarzschild-de Sitter black holes in four dimensions, the Lyapunov exponent of the nearly bound null geodesics is equal to the surface gravity \cite{Cardoso:2008bp}. In this section motivated by the relation to the inverted harmonic oscillator we treat photon sphere as a part of the holographic dual for an astrophysical black hole and in particular we identify the photon sphere as an effective holographic horizon and study the interpretation of quantum bounds.

Along the direction $x_1$ we obtain the inverted quantum harmonic oscillator \eq{wave1}, where we can identify an effective temperature for the quantum system that saturates a  bound of the form\footnote{One may examine a more general bound $\l\simeq c T_{eff}$ where $c\sim \cO(1)$ and depends on the scales of the theory.}
\be\la{bound1}
\l\le 2 \pi T_{eff}~.
\ee
We argue that the photon ring plays the role of an holographic horizon for these astrophysical black holes. In particular the photon ring is considered as an effective holographic horizon where the thermal bound on chaos for the dual quantum systems with a large number of degrees of freedom is saturated.

For the neutral static Schwarzschild black holes using Eq. \eq{lschw} we saturate the aforementioned bound for
\be\la{teffs}
\l^2=4 \pi^2 T_{eff}^2\Rightarrow T_{eff}=\ff{4}{3 \sqrt{3}} T~,
\ee
where $T$ is the Hawking temperature for the black hole and $\k_{Schw}= 2 \pi T$.
The temperature $T_{eff}$, has a physical meaning and turns out to be equal to the Unruh temperature of the system, as can be seen in the Appendix \ref{appA}. The photon ring can be approximated with a Rindler horizon where from its acceleration component the effective temperature can be read. In particular, the moving observer with a detector measures the number of quanta following the Bose-Einstein distribution with a temperature $T_{eff}$ \cite{Raffaelli:2021gzh}. Therefore the Unruh temperature associated to the photon ring is the one that saturates the bound \eq{bound1} for the Lyapunov exponent characterizing the unstable circular geodesics of the photon sphere. Accordingly, we can think of the photon ring as playing the role of an effective holographic horizon that determines the dual quantum theory temperature. 

For stationary black holes the bound has a richer  structure.  Let us consider a slowly rotating Kerr black hole where the Lyapunov exponent saturates a bound \eq{bound1}   on the photon sphere for the dual quantum systems with a large number of degrees of freedom.  The dependence $\l(\o,T)$ is in general non-linear and  non-polynomial, however for slowly rotating black holes, the effective temperature can be written as
\be
\la{teffk01}
T_{eff}\simeq \ff{T_{eff_0}}{\prt{1-\tilde{v}}^p}~,
\ee
where $\tilde{v}\ll 1$ and $p$ is an appropriate power. Using \eq{kerrl} the bound can be saturated  analytically with
\be \la{teffk1}
T_{eff_0}\simeq \ff{4 T}{3 \sqrt{3}}~,\qquad  \tilde{v}\simeq \ff{304}{27 p} \pi^2 a^2 T^2~,
\ee
where $T$ is the black hole temperature and $T_{eff_0}$ is equal to the Unruh temperature for the static case as can be seen by Eq. \eq{teffs}.  In presence of rotation the Lyapunov exponent increases by a factor that depends on the angular momentum. The form and the increase of the exponent for the outer ring, is in agreement with the variety of modifications of the suggested MSS bound  for thermal quantum systems with a large number of degrees of freedom \cite{Maldacena:2015waa}, in rotating environments \cite{Jahnke:2019gxr,Banerjee:2019vff,Djukic:2023dgk}, providing encouraging evidence of the quantum system correspondence and the holography of photon rings. Nevertheless, notice that in the extremal limit, the outer ring does not satisfy the Lyapunov bound in the traditional form and the effective temperature required in this case is independent of the black hole temperature, while in the inner ring the bound is trivially saturated.

Finally for Kerr-Newman black holes where angular momentum and charge are present the effective temperature increases further. In  the large mass limit,  using \eq{knl} we obtain
\be\la{teffkn}
T_{eff_0}\simeq \ff{4}{3 \sqrt{3}}T~,\qquad \tilde{v}\simeq \ff{16 \pi^2 \prt{19 a^2+6 q^2}T^2}{27 p}~,
\ee
where $T_{eff_0}$ is the Unruh temperature of the static system, and $T$ is the temperature of the black hole. Our findings suggest that the angular momentum and the electric charge, both act as enhancements of chaotic effects, implying a modification of the chaos upper bounds to higher values that depend of the charge and angular frequency.

When more than one parameters needed to characterize the black hole, the Lyapunov exponent is not always bounded by the surface gravity in the most generic case as we have seen. We suggest an alternative robust upper bound which is universally satisfied with the central quantity being the entropy of the black hole. The bound is motivated by the standard relation between the surface gravity and the entropy in Schwarzschild black holes $\k_{Schw}^2= \pi/(4S)$, and reads
\be \la{bound2}
\l^2 \le \ff{\pi}{4 S}~.
\ee
We have confirmed that is satisfied for the outer photon rings of the static or stationary black holes under study, for the whole parametric space including the extremal limits. This is a stronger bound which takes into account the extra charges and angular momenta. It would be interesting to examine its validity in other black holes as well as in thermal quantum systems.

\subsection{The (Inverted) Harmonic Oscillator}

In this section we study the behavior of the Lyapunov exponents by looking at the wave equations and relating them to the the one-dimensional (inverted) quantum harmonic oscillator.

We first briefly review some recent findings about the OTOC, which for a time-independent natural Hamiltonian $H$ is defined as
\be
C_T(t)=-\vev{[x(t),p(0)]^2}~,\quad \vev{\cO}=\ff{\tr[e^{-\b H} \cO]}{\tr e^{-\beta H}}~,
\ee
where the expectation value is taken over a thermal average with temperature $T=\beta^{-1}$. The OTOC  is rewritten in terms of the energy eigenstates $H\ket{n}=E_n \ket{n}$, as \cite{Hashimoto:2017oit}
\be
C_T(t)=\ff{1}{Z}\sum_n e^{-\b E_n} y_{2,nn}(t)~,
\ee
where we have defined
\be \la{motoc}
y_{2,nn}(t):=- \bra{n}[x(t),p]^2\ket{n}=\sum_m y_{nm}y_{nm}^*~,\qquad y_{nm}(t):= -i \bra{n}[x(t),p]\ket{m}~.
\ee
For a natural Hamiltonian we can compute the OTOC  by substituting $x(t)=e^{iHt}xe^{-iHt}$ to obtain
\be\la{ys}
y_{nm}(t)=\ff{1}{2} \sum_k x_{nk} x_{km} \prt{E_{km}e^{i E_{nk} t}-E_{nk} e^{i E_{km}t}} ~,
\ee
where $x_{nm}:=\bra{n}x\ket{m}$   and  $E_{nm}:=E_n-E_m$.
Therefore the information needed to determine $C_T(t)$ for the system is the energy spectrum and the matrix $x$. Let us consider the Hamiltonians
\be \la{ham1}
H_s=\ff{p^2}{ 2 m} + s\ff{m \o^2 x^2}{2},\quad s^2:=1~.
\ee
which correspond to the inverted ($s=-1$) harmonic oscillator and the simple harmonic oscillator ($s= 1$), respectively, and they are related to each other by
the analytic continuation $\o\to i \o$.
The associated quantum  harmonic oscillator equation reads
\be
-\ff{1}{2m}\psi''(x)+\ff{1}{2} s m \o^2 x^2 \psi(x)= E_s \psi(x)~,
\ee
and when compared to \eq{wave1}, we have
\be \la{harmo}
A_{ij}= -\frac{sm^2\o^2}{p_v^2}\delta_{ij}.
\ee
If $s=1$  the energy spectrum is given by $E_n=\prt{n+\ff{1}{2}} \o$ while the eq.\eq{ys} and \eq{motoc} give for the microcanonical OTOC $y_{2,nn}(t)$
\be
y_{2,nn}(t)=\d_{nm} \cos\o t~,\qquad  y_{nn}(t)=\cos^2 \o t~.
\ee
Therefore, the OTOC is periodic and independent of the energy and the temperature of the system \cite{Hashimoto:2016wme}
\be \la{so1}
C_T(t)=\cos^2\o t~.
\ee
For the inverted harmonic oscillator,  we could analytically continue the expression  \eq{so1} to get
\be
C_T(t)= \cosh^2(\o t)\simeq e^{2 \o t}~,
\ee
where the last equation is the leading contribution in the large time limit. Under this assumption the OTOC grows exponentially. Nevertheless, there is the problem  of imaginary energy, which makes a strict definition of the OTOC ambiguous. Moreover, the potential that corresponds to our system is unbounded from below and the definition and the assignment of a  temperature in the quantum model is problematic.

At any rate, the Penrose limit is the leading approximation on the photon ring physics and higher order terms could bound the potential from below. In this case, it has been proposed a simpler, more generic bound of the form $\l\le c T$, where $c=\cO(1)$ \cite{Hashimoto:2020xfr}. The bound is in direct agreement with our findings for the static neutral systems. For the rotating \eq{teffk1}, and charged black holes \eq{teffkn}, our results imply that when extra scales are present in the theory, the bound is still correct when one scale of the theory dominates the rest. The photon ring analysis suggests that, for massive black holes, the bound needs to be modified such that the parameter $c$ becomes a function of the rest of dimensionful scales of the theory, i.e., $c=c(a,q)$.

Finally, despite the aforementioned ambiguities we could follow an alternative way to make a direct analogy of the wave equation \eq{wave1} and the inverted harmonic oscillator from \eq{ham1} to obtain $A_{11} p_v^2=\l^2$. By accepting the existence of a bound $\l\le 2\pi T_q$, where $T_q$ is the temperature of the quantum system we obtain
\be
T_{q}=\ff{4}{3 \sqrt{3}} T~,
\ee
which is equal to the temperature $T_{eff}$ \eq{teffs} and the appropriate Unruh temperature for this system. As a side note, the $\sqrt{3}$ factor can be also understood as the redshift factor to the temperature on the ring
\be
V=\sqrt{-g_{tt}}=\ff{\D}{r^2+a^2}\bigg|_{r_0}=\ff{1}{\sqrt{3}}~.
\ee
For the rotating \eq{teffk1}, and charged black holes \eq{teffkn}, our results suggest an enhancement of the Lyapunov exponent and of the temperature $T_{eff}$ that saturates bound when extra scales are present in agreement with several independent observation following alternative formalisms \cite{Jahnke:2019gxr,Banerjee:2019vff,Djukic:2023dgk}.

 As a side comment we notice that a similar approach presented above for the OTOC can be followed for all the indicators of the quantum chaos and operator growth in relation to our systems using Eq. \eq{harmo}, for example, for the Krylov state and operator complexity \cite{Parker:2018yvk,Dymarsky:2021bjq,Hashimoto:2020xfr,Balasubramanian:2022tpr,Rabinovici:2022beu}.  It is known that the Krylov operator complexity \cite{Parker:2018yvk,Hashimoto:2023swv} for Gaussian operators in the integrable inverted harmonic operator grows exponentially as well, with a grown coefficient that is related to the frequency of the inverted harmonic operator. Through the relation of the photon ring geodesics to the quantum system, all the observations related to complexity measures in quantum systems, can be translated with our photon ring systems in a similar way we have done with the OTOC. We leave a deeper study about the OTOCs and other complexity measures, and their implications on photon ring for the near future.

\section{Discussions}

In this work, we present a generic formalism for the construction of the Penrose plane wave limit of general stationary axisymmetric spacetimes with a separable Hamilton-Jacobi equation of geodesics. The overall procedure boils down to finding the Killing-Yano tensor for black hole spacetimes, since the parallel-transported tetrad of covectors along the null photon ring geodesics is constructed from the Killing-Yano tensor, the geodesic's 4-velocity, and the covectors that lie in the plane transverse to the two existing ones. The class of spacetimes considered includes a wide variety of static or stationary black holes to which our construction directly applies.

We justify in generality a correspondence between the frequency matrix of the massless scalar equation and the quasinormal modes in Penrose limit, and the null geodesics deviation, which also suggested in \cite{Fransen:2023eqj}.  The massless wave equation in the Penrose plane wave has a frequency matrix that is straightforwardly related to the frequency matrix of the null congruences. The deviation of the null geodesics is identical in the original geometry and the plane wave limit, preserving covariantly the information. Therefore, it naturally follows that the wave equation in the Penrose limit is related to the deviation of null geodesics. In particular, the Lyapunov exponents characterizing the exponential divergence of the asymptotically bound geodesics in the near photon ring regime correspond in the Penrose limit to  the quasinormal modes in the large frequency limits.

Our universal formulas, \eq{lyapan0}, \eq{lyapa0}, for the Lyapunov exponents, the quasinormal modes and the angular frequencies, are directly applicable for a wide class of spacetimes. In particular the full analysis of the correspondence between the quasinormal modes and the properties of photon ring geodesics, boils down to producing the Penrose limit of the original spacetime black hole metric. Once this is done, all the details are directly readable from the plane wave metric.

Several fascinating consequences of the correspondence follow directly. Spacetimes that are solutions to the vacuum Einstein equations in the absence of null fluxes, have  Ricci flat Penrose limits. The matrix $A$  of the Penrose limit is related to the curvature tensor of the parallel frame along the photon ring geodesics and is traceless. This implies that the transformed Schrödinger equations of the massless wave equations, corresponding to the radial and angular geodesic deviations, always take the form of two "mirroring" harmonic oscillators. One is a normal harmonic oscillator along the stable (polar) direction, and the other is an inverted harmonic oscillator along the (radial) unstable direction, differing by a sign  in the frequency term. The unstable direction corresponds to the deviation of the null geodesics of the photon ring, while the stable one corresponds to the orbital frequency of the geodesics. The validity of the above statement is independent of the specific form of the metric as long as the spacetime is a solution to the vacuum Einstein equations, it follows directly from the Penrose limit plane wave approach.

Moreover, we relate the system of photon ring geodesics to a quantum inverted harmonic oscillator, with a potential identified by the Penrose limit metric. By making the mild assumption that the holographic principle applies to asymptotically flat axisymmetric spacetimes, and by arguing that the photon ring acts an effective holographic horizon of dual quantum systems with a large number of degrees of freedom,  we can motivate the study of possible bounds in frequencies, Lyapunov exponents and quasinormal modes by quantum physics.  The plane wave metric and the Lyapunov exponent can be expressed through the surface gravity of the photon ring of the black hole in a meaningful pattern. For example, in four-dimensional near extremal Schwarzschild-de Sitter black holes, the Lyapunov exponent of the nearly bound null geodesics is equal to the surface gravity \cite{Cardoso:2008bp}. Here, we generalize such relations with the surface gravity \eq{surf1} for generic spacetimes that include angular momentum and charge. We focus on the limit of large mass, where one scale dominates the rest, and we point out how such relations and bounds are modified in the presence of angular frequencies. In particular, by imposing the upper bounds \eq{bound1}, we analyze the effective temperatures that saturates them. For the case of the Schwarzschild black hole, this temperature is equal to the Unruh temperature of the near-photon ring regime, and consequently the photon ring acts as an effective holographic horizon for the thermal bound on chaos for the dual quantum systems with a large number of degrees of freedom. For rotating and charged black holes, our results  \eq{teffkn} suggest an enhancement of the bound and the effective temperature when extra scales are present, in agreement with several independent observations about quantum chaos when similar extra charges are present. These are strong indications that the photon ring can be considered as an effective holographic horizon for a wide range of asymptotically flat axisymmetric spacetimes with dual quantum systems of a large number of degrees of freedom. In general, having the photon ring play a crucial role as part of the holographic duality is in agreement with other independent recent approaches for astrophysical black holes \cite{Hadar:2022xag} and AdS black holes \cite{Hashimoto:2023buz,Riojas:2023pew,Dodelson:2023nnr}.

Additionally, we suggest an a alternative stronger bound, where the Lyapunov exponent is bounded by the inverse of entropy \eq{bound2} instead of the surface gravity. We find that the bound is universally satisfied for the photon rings considered in this section for the  whole parametric space of black hole parameters including the extremal limits. It would be interesting to study the validity of this new bound in quantum systems.

The existence of a generic framework allows us to investigate possible universalities in the Lyapunov exponents. For example, for the Kerr black hole, we find that for the outer photon ring, the orbital frequency is lower than the Lyapunov exponent, with implication that carry on to the inequality between the real and imaginary levels of the quasinormal mode spectrum. Moreover, in the class of spacetimes we consider the Hamilton-Jacobi equations are separable and a conserved Carter constant exists. It would be interesting to investigate limitations of generality of our correspondence in black hole spacetimes with less symmetry. Additionally, our findings suggest that the various perturbations in the Kerr-Newman black hole can be separated when employing the Penrose limit plane wave metric. This separation and the resulting equivalence to the massless wave equation occurs similarly to other black holes examined here. This presents an alternative approach to circumvent the longstanding issue of nonseparability in the perturbation equations for these rotating charged black holes \cite{Pani:2013wsa,Li:2021zct}.
Furthermore, it is known that the Seiberg-Witten curves can be used to compute analytically the spectrum of quasinormal modes \cite{Aminov:2020yma,Bonelli:2021uvf,Bianchi:2021xpr,Bianchi:2021mft}. It is worthy to delve deeper into possible connections and explore how our techniques and these approaches can be amenable.

Furthermore, our work provides generic formulas for the eikonal quasinormal modes and the null geodesics across a wide spectrum of spacetimes, offering potential direct implications in astrophysics. The curves observed in the shadow images of black holes are linked to the impact parameter of spherical photon orbits around these black holes. Consequently, the eikonal quasinormal modes are intimately connected to the shadows cast by black holes \cite{Stefanov:2010xz,Jusufi:2019ltj,Cuadros-Melgar:2020kqn}. Any deviation from the expected correspondence between the Lyapunov exponents and the corresponding quasinormal modes might indicate physics beyond  general relativity \cite{Konoplya:2017wot,Glampedakis:2019dqh,Chen:2022ynz,Konoplya:2022gjp}.  Exploring how the broad applicability of the correspondence in the Penrose limit can contribute to such studies would be of considerable interest.
\section*{Acknowledgments }
The research work of DG is supported by the National Science and Technology Council (NSTC) of Taiwan with the Young Scholar Columbus Fellowship grant 113-2636-M-110-006.

\appendix
\section{Appendix: A Metric That Admits Killing-Yano Tensor}
\la{appA0}
In this short appendix we present for convenience the form of a metric that admits a Killing-Yano tensor with the choices for the functions $A_i$ and $B_i$ given by \eq{ansatzais} and \eq{ansatzbis} for the special solution \eq{cond}. The metric \eq{metric1} reads
\be \nn
g_{\mn}(r,\theta) =
\left[\begin{array}{cccc}
-\ff{ \D(r)-a^2 s_\th^2}{ r^2+a^2 c_{\th}^2} & 0 & \ff{ a\prt{\D(r)-r^2-a^2}s_\th^2}{ r^2+a^2 c_{\th}^2} & 0\\
0 & r^2+a^2 c_\th^2 & 0 & 0\\
\ff{ a\prt{\D(r)-r^2-a^2} s_\th^2}{ r^2+a^2 c_{\th}^2} & 0 & \ff{\prt{\prt{a^2+r^2}^2-a^2 \D(r) s_\th^2}s_{\th}^2}{r^2+a^2 c_{\th}^2} & 0\\
0 & 0 & 0 & \ff{r^2+a^2 c_{\th}^2}{\D(r)}
\end{array}\right].
\ee

\section{Appendix: Unruh Temperature of the Schwarzschild Ring}
\la{appA}
In this appendix we show that the effective temperature \eq{teffs}, that comes up by requiring the satisfaction of a quantum-like bound for the spherically symmetric static black hole, matches with the Unruh temperature of an effective Rindler horizon on the photon sphere. We review the derivation of the Unruh temperature on the photon sphere geodesics. The approach follows \cite{Alsing:2004ig,Raffaelli:2021gzh}. The near-photon sphere limit geometry approaches effectively the Rindler spacetime
\be
ds^2=- (\a x)^2 r^2 dt^2+dx^2~,
\ee
where $\a$ is the proper acceleration related to the effective potential of the geodesic \cite{Cardoso:2008bp} and equal to
\be \a=\ff{1}{2}\ff{\sqrt{g_{tt}(r_0)\prt{4 g_{tt}(r_0)-2 r_0^2 g_{tt}''(r_0)}}}{r_0}~.
\ee
The frequency spectrum $S(\o)$ considering the relativistic Doppler effect for an accelerated observer along $x$ is given by
\be
S(\o)=\prtt{\int_{-\infty}^\infty d\t e^{i \o \t}e^{i\varphi(\t)}}^2=
\ff{2\pi}{\o \a} \ff{1}{e^{\ff{2\pi\o}{\a}}-1}~,
\ee
where
\be
\varphi:= \ff{\o}{\a} e^{\a \t}~.
\ee
From the Planck factor one reads as usual the Unruh temperature
\be
T_{Unruh}=\ff{\a}{2\pi}=\ff{4 T}{3 \sqrt{3}}~,
\ee
which turns out to be equal to $T_{eff}$ from \eq{teffs}.

\bibliographystyle{JHEP}

\end{document}